\documentclass[
 reprint,
 amsmath,amssymb,
 aps,
]{revtex4-2}
\usepackage[dvipsnames]{xcolor}
\usepackage{graphicx}
\usepackage{dcolumn}
\usepackage{bm}
\usepackage{braket}
\usepackage{svg}
\usepackage{hyperref}
\usepackage[T1]{fontenc}
\usepackage{notes2bib}

\newcommand{\ISb}[1]{\makebox[0pt]{{#1}}} 
\newcommand{\ISc}[1]{\makebox[0pt]{\raisebox{-0.5\height}{#1}}}
\newcommand{\ISt}[1]{\makebox[0pt]{\raisebox{-\height}{#1}}}

\newcommand{\tpmhz}{\ensuremath{~2\pi\text{MHz}}}

\newcommand{\sftl}[1]{\makebox[0pt][l]{\hspace*{0.8mm}\raisebox{-3.2mm}{#1}}} 
\newcommand{\sftr}[1]{\makebox[0pt][l]{\hspace*{-6.0mm}\raisebox{-3.2mm}{#1}}} 

\DeclareUnicodeCharacter{2212}{-}

\begin{document}


\title{Cavity QED systems for steady-state sources of Wigner-negative light}

\author{Alex Elliott\textsuperscript{1,2}}
\email{alexander.elliott@auckland.ac.nz}
\author{Scott Parkins\textsuperscript{1,2}}%
\email{s.parkins@auckland.ac.nz}
\affiliation{%
\textsuperscript{1}The Dodd-Walls Centre for Photonic and Quantum Technologies, Auckland, New Zealand\\
\textsuperscript{2}Department of Physics, University of Auckland, Auckland, New Zealand.
}%

\date{\today}

\begin{abstract}
We present a theoretical investigation of optical cavity QED systems, as described by the driven, open Jaynes-Cummings model and some of its variants, as potential sources of steady-state Wigner-negative light. We consider temporal modes in the continuous output field from the cavity and demonstrate pronounced negativity in their Wigner distributions for experimentally-relevant parameter regimes. 
We consider models of both single and collective atomic spin systems, and find a rich structure of Wigner-distribution negativity as the  spin size is varied. 
We also demonstrate an effective realization of all of the models considered using just a single ${}^{87}$Rb atom and based upon combinations of laser- and laser-plus-cavity-driven Raman transitions between magnetic sublevels in a single ground hyperfine state.
\end{abstract}

\pacs{Valid PACS appear here}
\maketitle


\section{Introduction}
    Experiments in optical cavity quantum electrodynamics (cavity QED), underpinned on the theoretical side by the celebrated Jaynes-Cummings model, have been a leading foundation for the generation of nonclassical states of light \cite{Miller2005}. 
    This a tribute to the efforts of experimentalists to achieve the most optimal operating conditions possible, by (amongst other things) controlling the motion and number of atoms, minimizing cavity losses, and maximizing atom-cavity coupling strengths. 
    In fact, modern optical cavity QED experiments are capable of remarkable coupling strengths between single atoms and single photons (see, for example, \cite{Hood1998,Hood2000,Gehr2010,Samutpraphoot2020,Urunuela2022}), which offer exciting possibilities for the manipulation of quantum states of both atoms and light. 
    
    With regards to the continuous, or steady-state generation of nonclassical light in optical cavity QED, the greatest focus has arguably been on antibunched light \cite{McKeever2003,Birnbaum2005,Dayan2008,Aoki2009,Gallego2018}, associated with two-level-type behavior in atoms, and squeezing \cite{Ourjoumtsev2011}, associated with correlated photon-pair emission. Another, more general class of nonclassical light is yet to be explored in this context though -- so-called ``Wigner-negative'' states of light. These are states of a mode of the electromagnetic field for which the associated Wigner distribution takes on negative values. In the steady-state emission from a cavity QED system, they will correspond to specified, temporal modes of the cavity output field.
    
    Production of Wigner-negative states of light is of considerable current interest in quantum optics, particularly as these states constitute a key resource in the pursuit of optical quantum information technologies \cite{Mari2012,Veitch2012,Veitch2013}. A substantial research effort has, in particular, targeted the generation of Schr\"odinger-cat-like states and Fock states. However, despite significant advances in the precise control of quantum systems, consistent and deterministic, steady-state production of Wigner-negative light remains challenging.
    
    A variety of mechanisms have been proposed and implemented for the generation of Wigner-negative states. Conditional schemes can probabilistically prepare a variety of such states, heralded by the detection of a photon, or photons, separated from the original field by a beamsplitter \cite{Ourjoumtsev2006,Gerrits2010,Huang2015,Baune2017,Takase2022,Konno2024}. Alternatively, pulses of Wigner-negative light can be generated on-demand from the transient evolution of a quantum system initially prepared in a specific state \cite{Brown2003,Gonzalez2015,Hacker2019,Paulisch2019,Groiseau2021}. Finally, homodyne phase-controlled feedback has also been suggested as a means for the steady-state generation of Wigner negativity from highly nonlinear optical media \cite{Joana2016,Li2021}. 
    
    Recently, though, it has been suggested, and demonstrated, that Wigner-negative states can be produced from the nonlinearity of a driven, one-dimensional two-level quantum emitter in temporal modes of the continuous outgoing field \cite{Quijandria2018,Strandberg2019,Lu2021}. Further work has suggested that this may be a promising and robust mechanism to generate steady-state Wigner-negative light from a variety of other quantum systems \cite{Strandberg2021,Kleinbeck2023}.
    
    Spurred on by these findings, in this work we expand the investigation of Wigner-negative states to temporal modes 
    in the steady-state output field from optical cavity QED systems; in particular, to the Jaynes-Cummings model and its variants, such as the multi-atom Tavis-Cummings model. We demonstrate that, in appropriate (and experimentally relevant) regimes, the open Jaynes-Cummings model can reproduce the kind of Wigner-negative distributions found in \cite{Quijandria2018,Strandberg2019,Lu2021}, but also that its multi-atom variants open up a family of new possibilities related to the multiple-excitation opportunities they afford.   
    
    Further to this, and encouraged by the aforementioned experimental realizations of exceptional atom-field coupling strengths in the optical regime of cavity QED, we then present results for a full-atomic-structure model of a single ${}^{87}$Rb atom coupled to a quantized cavity mode and driven by auxiliary laser fields in such a way as to realize, effectively, the dynamics described by both the Jaynes-Cummings model and its higher-spin variants (e.g., the Tavis-Cummings and Dicke models). We do this by making use of the multilevel structure of the atom and laser-plus-cavity driven Raman transitions between ground-state magnetic sublevels. This comprehensive model of a real atom coupled to a cavity mode is shown to be capable of successfully reproducing the multitude of new Wigner-negative distributions presented earlier for the idealized models. 

    To summarize the contents of this work, in Section II an overview of the physical setup that we consider, and the theoretical modeling that we use, is presented. This includes a specification of the temporal modes that we choose to study, which are based upon matching the mode profile to the emission spectrum of the cavity mode, and the definition of, and means by which we calculate, the Wigner distributions for temporal mode states of the cavity output field. 
    
    In Section III these choices and methods are applied to the case of one or a few two-level atoms, as described by the driven and open Jaynes-Cummings and Tavis-Cummings models. These investigations help us to understand the influence of cavity loss and atomic spontaneous emission on the generation of Wigner-negative states. In particular, they reveal strong sensitivity to spontaneous emission as the number of two-level atoms increases, which prompts us to consider an alternative, effective realization of the collective multi-atom models based upon Raman transitions between ground-state magnetic sublevels in a single alkali atom. This approach avoids spontaneous emission, which justifies the idealized models that are subsequently shown to exhibit a rich variety of Wigner-negative states, resulting from more intricate photon superposition states, as the total effective spin is increased. 
    
    Finally, in Section V we present results for a full-atomic-structure model of a single ${}^{87}$Rb atom coupled to a cavity mode and to combinations of laser fields in configurations that are carefully tailored to produce effective dynamics matching those of the idealized models considered earlier. These results demonstrate the feasibility of producing Wigner-negative states in temporal modes of the cavity output field for parameters in the strong-coupling regime of optical cavity QED.

\section{Cavity QED System}

\subsection{Physical set up}

\begin{figure}[htp]
        \centering
\begingroup%
  \makeatletter%
  \providecommand\color[2][]{%
    \errmessage{(Inkscape) Color is used for the text in Inkscape, but the package 'color.sty' is not loaded}%
    \renewcommand\color[2][]{}%
  }%
  \providecommand\transparent[1]{%
    \errmessage{(Inkscape) Transparency is used (non-zero) for the text in Inkscape, but the package 'transparent.sty' is not loaded}%
    \renewcommand\transparent[1]{}%
  }%
  \providecommand\rotatebox[2]{#2}%
  \newcommand*\fsize{\dimexpr\f@size pt\relax}%
  \newcommand*\lineheight[1]{\fontsize{\fsize}{#1\fsize}\selectfont}%
  \ifx\svgwidth\undefined%
    \setlength{\unitlength}{219.82391167bp}%
    \ifx\svgscale\undefined%
      \relax%
    \else%
      \setlength{\unitlength}{\unitlength * \real{\svgscale}}%
    \fi%
  \else%
    \setlength{\unitlength}{\svgwidth}%
  \fi%
  \global\let\svgwidth\undefined%
  \global\let\svgscale\undefined%
  \makeatother%
  \begin{picture}(1,0.30653577)%
    \lineheight{1}%
    \setlength\tabcolsep{0pt}%
    \put(0.37486139,0.00565346){\color[rgb]{0,0,0}\makebox(0,0)[t]{\lineheight{1.25}\smash{\begin{tabular}[t]{c}\ISt{$\kappa$}\end{tabular}}}}%
    \put(0,0){\includegraphics[width=\unitlength,page=1]{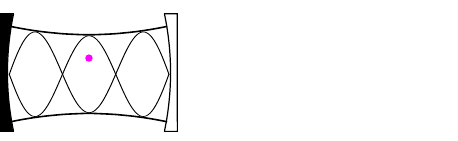}}%
    \put(0.21334209,0.27853031){\color[rgb]{0,0,0}\makebox(0,0)[t]{\lineheight{1.25}\smash{\begin{tabular}[t]{c}\ISb{$\gamma$}\end{tabular}}}}%
    \put(0.1942127,0.12754433){\color[rgb]{0,0,0}\makebox(0,0)[t]{\lineheight{1.25}\smash{\begin{tabular}[t]{c}\ISc{$g$}\end{tabular}}}}%
    \put(0,0){\includegraphics[width=\unitlength,page=2]{BasicCavitySchematicNew.pdf}}%
    \put(0.45515791,0.1736143){\makebox(0,0)[t]{\lineheight{1.25}\smash{\begin{tabular}[t]{c}\ISb{$\hat{a}_\text{out}$}\end{tabular}}}}%
    \put(0.55271677,0.27918595){\makebox(0,0)[t]{\lineheight{1.25}\smash{\begin{tabular}[t]{c}\ISb{$\mathcal{E}_\text{lo}(t)$}\end{tabular}}}}%
    \put(0.09171222,0.00657993){\makebox(0,0)[t]{\lineheight{1.25}\smash{\begin{tabular}[t]{c}\ISt{$\Omega$}\end{tabular}}}}%
    \put(0,0){\includegraphics[width=\unitlength,page=3]{BasicCavitySchematicNew.pdf}}%
  \end{picture}%
\endgroup%

        \caption{Schematic of a cavity QED system, with a resonant atom-cavity coupling strength $g$, atomic spontaneous emission linewidth $\gamma$, and cavity field decay rate $\kappa$. The system can be pumped by a coherent laser field driving the cavity mode with a strength $\mathcal{E}$, or by auxiliary lasers directly driving the atom, with Rabi frequencies $\Omega$. Temporal modes of the output field can be measured through balanced homodyne detection with a time-dependent local oscillator.}
        \label{fig:1}
    \end{figure}

    The models explored in this work are based on a cavity QED set-up as depicted in Fig. \ref{fig:1}. An atom is coupled to a single mode of a high-quality optical resonator, with a (maximum) resonant coupling strength $g$. The system can be driven by directly illuminating the atom, or by coherently pumping the cavity mode. The atom has a spontaneous emission linewidth $\gamma$, and the cavity field decays due to photon losses through the transmitting mirror at a rate $\kappa$. 
    
        The driving fields are all strictly classical; quantum features in the output field are necessarily due to the atom-field interactions within the cavity. However, the cavity mode itself does not demonstrate Wigner-negativity in the regimes we explore. Instead, operating regimes are chosen to enhance quantum features of the cavity output field. One approach is for the systems to operate in a quasi-bad-cavity limit, where the cavity field decay rate is larger than the effective rate of coupling light from the atom. Any light transferred to the cavity mode will then rapidly escape into the output field, preserving features of the interaction. A second avenue makes use of the anharmonic Jaynes-Cummings ladder in a strongly-coupled regime. The dressed atom-cavity states give a photon blockade effect, such that the overall system behaves effectively as a one-dimensional, two-level emitter \cite{TianCarmichael1992}. 
        
        In order that characteristics of the interaction can be detected as Wigner-negativity in the output field, it is important that the systems we consider possess a large single-atom cooperativity, as specified by the parameter $C = 2g^2/\kappa\gamma$. In particular, we require $C\gg 1$, which ensures a strong atom-field interaction occurs before quanta are lost, and hence that the cavity emission is strongly influenced by the atom. The ratio of coupling and cavity decay rates, $g/\kappa$, is fixed when choosing the operating regime to optimize cavity emission properties. It is therefore incumbent on the ratio $g/\gamma$ to be sufficiently large that a suitable cooperativity is achieved.
        
        The cavity needs also be manufactured in such a way that there is a single dominant photonic decay channel, to preserve features of the interaction. This can be achieved by having one ``bad'' cavity mirror, which provides the dominant decay channel for the cavity field.

\subsection{Theoretical modeling}

\subsubsection{Master equation}
 The cavity QED system is described by a density operator for the composite quantum state, $\hat{\rho}$. Evolution of the density operator is governed by a Lindblad master equation of the general form ($\hbar$=1)
    \begin{equation}
    \dot{\hat{\rho}} \equiv\mathcal{L}\hat{\rho}= -i[\hat{H},\hat{\rho}] + \mathcal{D}_C\{\hat{\rho}\}+\mathcal{D}_A\{\hat{\rho}\},
    \end{equation}
    where $\mathcal{D}_C$ and $\mathcal{D}_A$ describe decay of the cavity mode field and spontaneous emission of atomic excited states, respectively. All the models to be considered can be written in an interaction picture, eliminating any explicit time-dependence of the Hamiltonian, $\hat{H}$.
  
\subsubsection{Temporal modes}
    The field propagating from a transmitting cavity mirror depends on the cavity mode according to the input-output relation \cite{Gardiner1985} 
    \begin{equation}
        \hat{a}_\text{out}(t) = \sqrt{2\kappa}\hat{a}(t)-\hat{a}_\text{in}(t),
    \end{equation}
    where $\hat{a}$ is the annihilation operation for the cavity field mode.
    In this work, the leaky-mirror input $\hat{a}_\text{in}(t)$ is taken to be a vacuum field.
    Temporal modes are defined in the travelling-wave output field by an envelope function, $f(t)$, such that
    \begin{equation}
        \hat{a}_f = \int_{-\infty}^\infty f(t)\hat{a}_\text{out}(t)dt,
    \end{equation}
    with normalisation 
    \begin{equation}
        \int_{-\infty}^\infty|f(t)|^2dt=1,
    \end{equation}
    which ensures the bosonic commutation relation $[\hat{a}_f^{},\hat{a}_f^\dagger] = 1$. 

     In steady state, light is continually emitted into the output field with a coherence time limited by decay rates in the system. If a temporal mode is to exhibit negativity in its Wigner distribution, the envelope function should generally reflect the time-scale of correlations in the atom-cavity system. The steady-state cavity field amplitude correlation function is a logical candidate to achieve this temporal mode-matching, with
    \begin{equation}
        f(t)\propto\braket{\hat{a}^\dagger(0)\hat{a}^{}(t)}_{\text{ss}} -\braket{\hat{a}^\dagger}_{\text{ss}}\braket{\hat{a}}_{\text{ss}}.
    \end{equation}
    The correlation function can be calculated from the master equation using the quantum regression theorem \cite{Carmichael1999}. In practice, a cut-off point is chosen where the relative magnitude of $f(t)$ is sufficiently small. Other choices of temporal mode function also work similarly well for the purpose of generating Wigner-negativity, so long as they have a similar temporal width to the correlation functions. In this work, the function $f(t)$ is chosen according to the above equation.
    
\subsubsection{Wigner function reconstruction}
    A Wigner distribution for the quantum state of a temporal mode can be defined with respect to the quadrature operators
    \begin{equation}
        \hat{X}_f = \hat{a}_f + \hat{a}_f^\dagger,~~~\hat{Y}_f = i\left(\hat{a}_f^\dagger-\hat{a}_f\right).
        \label{quadratures}
    \end{equation}
    The Wigner distribution is always non-negative for any classical field, but it is possible to generate quantum states which violate this condition. The magnitude of Wigner-negative nonclassicality of a state can be quantified by the total negative volume \cite{Kenfack2004},
    \begin{equation}
        \mathcal{N} = \frac{1}{2}\int \left( |W(x,y)|-W(x,y) \right) dxdy.
    \end{equation}
    Naturally, the Wigner distribution cannot be thought of as a genuine probability distribution if it is to permit negative values. However, the marginal distributions are true probability distributions for eigenvalues of quadrature amplitudes. This connection can be used to experimentally determine the Wigner distribution using quantum state tomography, through balanced homodyne detection \cite{Smithey1993,Ourjoumetsev2006}.
    
    The quantum state of a temporal mode can be determined by integrating the homodyne difference-current (Fig.~\ref{fig:1}), with a time-dependent local-oscillator amplitude matched to the temporal profile ($\mathcal{E}_\text{lo}\propto f(t)$). The measured distributions are then related to the underlying quantum state through maximum-likelihood estimation \cite{Lvovsky2004}. 
    
    While the tomography process can be simulated using quantum trajectory theory \cite{Carmichael2008}, in practice it is more convenient to perform numerical calculations using the input-output theory for quantum pulses \cite{Kiilerich2019}, which introduces a fictitious ``capture cavity'' in the output field from the cavity QED system. The composite system steady-state is found by solving the equation $\mathcal{L}\hat{\rho}=0$, either directly for reduced, effective models, or iteratively for the full atomic models we also consider later in this work. In this case, the temporal mode state is calculated by numerically integrating the time-dependent master equation. 
    
    All numerical calculations in this work are performed using QuTiP \cite{qutip1,qutip2,mee}. Cavity modes are simulated on a truncated Fock basis, and we choose the truncation at a sufficiently high level to ensure that our numerical results are accurate. Details about the implementation of the input-output method are given in Appendix A.

    Before continuing, we note that any coherent amplitude in the output field merely translates the Wigner distribution in phase space, without affecting the shape or negativity. In principle, this means that coherent driving of the cavity mode can be implemented through either mirror, without affecting the desirable features in the measured quantum state.

\section{Two-level Atoms}
    Remarkable advances in circuit QED have enabled strong coupling of effective two-level emitters to electromagnetic fields propagating in waveguides, thus realizing effectively one-dimensional atom-light systems. Using just such a system, it was demonstrated, first theoretically \cite{Quijandria2018,Strandberg2019} and then experimentally \cite{Lu2021}, that suitably chosen temporal modes in the steady-state emission from a coherently-driven two-level atom can exhibit pronounced negativity in their Wigner distributions. Theoretical studies have also demonstrated similar behavior for a chain of 1-D emitters \cite{Kleinbeck2023}. 

    The strong coupling of a two-level atom to an optical cavity mode can also, under appropriate conditions, lead to largely one-dimensional behavior, with the dominant emission channel from the coupled system being through one of the cavity mirrors. Inspired by the above-mentioned work in the microwave regime of circuit QED, in this section we start our investigation of Wigner-negative temporal modes in optical cavity QED by considering the (dissipative) Jaynes-Cummings model for a single two-level atom. As we shall see, this classic model and its variants are indeed still capable of turning up novel and topical forms of nonclassical light.

\subsection{A single two-level atom}
    A single two-level atom coupled to a cavity mode can be described by a Jaynes-Cummings model. Assuming the bare cavity mode and atomic resonance frequencies are equal, the Jaynes-Cummings Hamiltonian, with driving of the cavity mode by an external laser, takes the form (in a frame rotating at the laser frequency)
    \begin{equation}
        \hat{H}_{\rm JC} = \Delta \left(\hat{a}^\dagger\hat{a}^{} + \hat{\sigma}_z\right) + g(\hat{a}^\dagger\hat{\sigma}_- + \hat{\sigma}_+\hat{a}) + \mathcal{E}(\hat{a} + \hat{a}^\dagger), 
        \label{JCHamiltonian}
    \end{equation}
    where $\Delta$ and $\mathcal{E}$ are the detuning and  strength of the driving field, respectively, and $\hat{\sigma}_j$ ($j=\pm,z$) are the usual spin-1/2 atomic operators. Note that the driving of the system can also be implemented by direct excitation of the atom, where the third term in Hamiltonian (\ref{JCHamiltonian}) is replaced with $(\Omega /2)(\hat{\sigma}_++\hat{\sigma}_-)$ and $\Omega$ is the Rabi frequency. Results are essentially identical regardless of the driving mechanism; we present results for both forms of driving in this section. Finally, cavity field decay and atomic spontaneous emission are included, respectively, via the terms
   \begin{equation}
    \mathcal{D}_C\{\hat{\rho}\} = \kappa \left( 2\hat a\hat\rho\hat a^\dag - \hat a^\dag \hat a\hat\rho - \hat\rho\hat a^\dag \hat a \right) ,
    \end{equation}
    and
    \begin{equation}\label{eq:D_A}
    \mathcal{D}_A\{\hat{\rho}\} = \frac{\gamma}{2} \left( 2\hat\sigma_-\hat\rho\hat\sigma_+ - \hat\sigma_+ \hat\sigma_-\hat\rho - \hat\rho\hat\sigma_+ \hat\sigma_- \right) .
    \end{equation}

    The Wigner-negative states demonstrated in the emission from a driven two-level atom are described approximately by superpositions, or more generally mixtures of the vacuum and single-photon states \cite{Quijandria2018,Strandberg2019,Lu2021}. Achieving this is evidently associated with successive photon emissions being sufficiently well-separated in time. 
    With this in mind, we note that a cavity QED system such as we are considering is well known to produce  antibunched photon counting statistics in two, distinct operating regimes. 
    
\begin{figure}[htp]
        \centering
        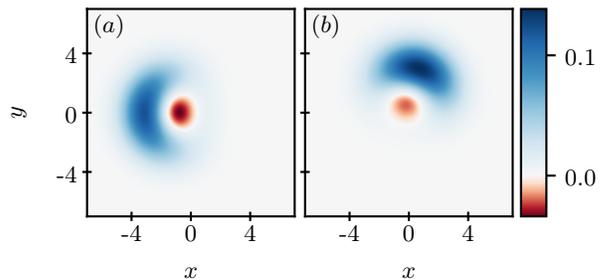
        \caption{Temporal-mode Wigner distributions in the output field of optical cavity QED systems in two distinct operating regimes: (a) in a quasi-bad-cavity regime, with parameters $\{g,~\kappa,~\gamma\} = \{100,~200,~6\}~2\pi\text{MHz}$ ($C = 16.7$), and $\Omega =0.75g $, and (b) in a strong-coupling regime,  with $\{g,~\kappa,~\gamma\} = \{100,~40,~6\}~2\pi\text{MHz}$ ($C=83.3$), and $\Omega = 0.4g $.}
        \label{TLAWigsrealparams}
    \end{figure}

    Firstly, in a quasi-bad-cavity limit, where $\kappa>g\gg \gamma$, the cavity mode primarily directs the atomic emission into the cavity output field, which adiabatically follows the atomic state. This regime of effectively one-dimensional atomic spontaneous emission is actually quite analogous to the semi-infinite-waveguide, circuit QED system, and so, perhaps unsurprisingly, exhibits a very similar Wigner-negative state for a suitably chosen temporal mode of the cavity output field, as shown in Fig.~\ref{TLAWigsrealparams}(a). However, it is notable that we consider realistic parameters for the regime of optical cavity QED, and we include free space atomic spontaneous emission in the model.
    
    Alternatively, in a strong-coupling limit where $g\gg\kappa,\gamma$, antibunching emerges from single-photon blockade within the nonlinear Jaynes-Cummings ladder of energy eigenstates \cite{TianCarmichael1992}. If the driving field is tuned such that $\Delta\simeq \pm g$, and $\kappa\gg\gamma$, then the composite atom-cavity system behaves effectively as a one-dimensional two-level emitter. As illustrated in Fig.~\ref{TLAWigsrealparams}(b), this does indeed also facilitate the generation of similar Wigner-negative states, once again for parameters that should be accessible to experiment.
    
    Light emitted from the cavity comprises a coherent and an incoherent component. The coherent component gives a finite coherent amplitude to the output field, and hence shifts the temporal-mode Wigner distribution in phase space. Displacing the state to remove this coherent amplitude reveals the incoherent component of the emission that arises from the presence of the atom and underlies the structure of the Wigner distribution. Doing so, we find, for Fig.~\ref{TLAWigsrealparams}(a), that the incoherent portion of the temporal mode state has zero, one, and two photon populations of $0.73$, $0.17$, and $0.09$, respectively. The atomic excited state probability is found to be $0.28$, which clearly limits the level of incoherent excitation in the temporal mode.
    Similar levels of excitation are observed in the photon-blockade temporal mode state. 
    
    Although the temporal mode states exhibit negativity in their Wigner distributions, this does not mean the states are pure. In fact, Tr$\{\hat{\rho}^2\} = \{0.76,~0.87\}$ for Figs.~\ref{TLAWigsrealparams}$\{(a),~(b)\}$, respectively, and the purities remain similar even in the absence of spontaneous emission. That the states are Wigner-negative but not pure is general to all the results presented in this work.
    
\begin{figure}[htp]
        \centering
        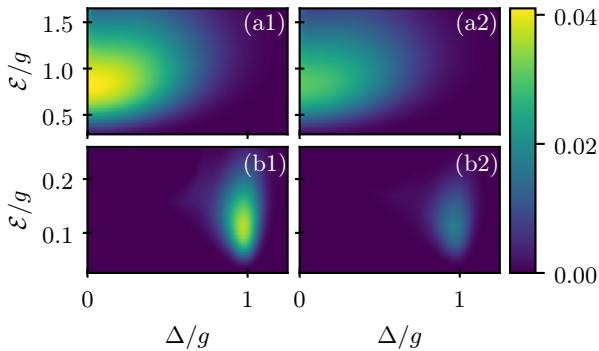
        \caption{Negative volume of the Wigner distributions in (a) the quasi-bad-cavity limit with $\kappa = 2g$, and (b) the single-photon blockade regime with $\kappa = 0.2g$, as a function of the strength and detuning of the laser driving the cavity, for atomic linewidths $(1)$ $\gamma = 0$ and $(2)$ $\gamma = 0.05g$ (corresponding to $C=\{20,~200\}$ for $\{(a2),~(b2)\}$).}
        \label{QBCOPBScan}
    \end{figure}

    The total negative volume shown in Fig.~\ref{QBCOPBScan} quantifies the magnitude of Wigner-negativity for the generated states over a range of strengths and detunings of the driving laser. This demonstrates the viability of generating Wigner-negative light for a range of parameters in the two operating regimes. The quasi-bad-cavity system performs optimally on resonance, where the cavity-directed emission is strongest, while the photon-blockade regime requires $\Delta \sim \pm g$ to isolate the single-photon resonance in the Jaynes-Cummings ladder.

\begin{figure}[htp]
        \centering
        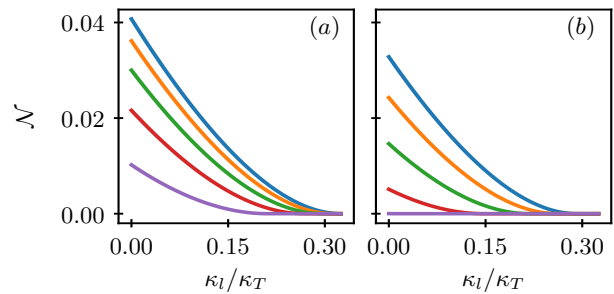
        \caption{Negative volume of the Wigner distribution as the relative contribution from parasitic cavity loss at rate $\kappa_l$ is increased while the total cavity decay rate remains constant in (a) the quasi-bad-cavity regime with $\{\kappa_T,~\mathcal{E}\}/g = \{2,~0.75\}$, and (b) the photon-blockade regime with $\{\kappa_T,~\mathcal{E}\}/g = \{0.2,~0.1\}$. The total negativity also decreases for increasing spontaneous emission rates; from the uppermost curve to the lowest curve, the spontaneous emission rate is given by $\gamma/g = \{0,~0.02,~0.05,~0.1,~0.2\}$, corresponding to  (a) $C= \{\infty,~50,~25,~10,~5\}$ and (b) $C= \{\infty,~500,~250,~100,~50\}$.}
        \label{QBCOPBCompare}
    \end{figure}

    While we already take into account free space atomic spontaneous emission, another practical aspect to consider is extraneous cavity loss. That is, in practice a cavity cannot be manufactured to be perfectly one-sided. Total cavity loss ($\kappa_T$) is a combination of emission into the (desirable) measured output channel ($\kappa$) and parasitic losses into other, non-monitored modes ($\kappa_l$).  Fig.~\ref{QBCOPBCompare} shows the effect that parasitic losses have on the ``quantumness'' of states in the measured output field, where the total cavity decay rate $\kappa_T = \kappa + \kappa_l$ is held constant, while the ratio $\kappa_l/\kappa_T$ is increased. This emphasizes the importance of engineering the emission to be dominated by a single output channel from the cavity.

\subsection{A collection of two-level atoms}
    The statistics of emission from a two-level system inherently limits the number of photons in the Wigner-negative superposition states. To explore the possibility of more exotic quantum-optical states, it is necessary to consider systems with multiphoton emission properties. A collection of $N$ two-level atoms coupled to a single cavity mode is therefore a logical place to start. Such a system is described by the multiatom version of the Jaynes-Cummings model, that is, the Tavis-Cummings model, with Hamiltonian
    \begin{equation}
        \hat{H}_{TC} = \frac{g}{\sqrt{N}}\sum_{n=1}^N (\hat{a}\hat{\sigma}_+^{(n)} +\hat{\sigma}_-^{(n)}\hat{a}^\dagger ) + \mathcal{E}(\hat{a} + \hat{a}^\dagger) ,
    \end{equation}
    which assumes identical coupling strengths for the atoms. Once again, we add a term to model coherent driving of the cavity field, and a spontaneous emission term of the form (\ref{eq:D_A}) is added to the master equation for each atom (i.e., we assume independent-atom spontaneous emission).
    
\begin{figure}[t]
        \centering
        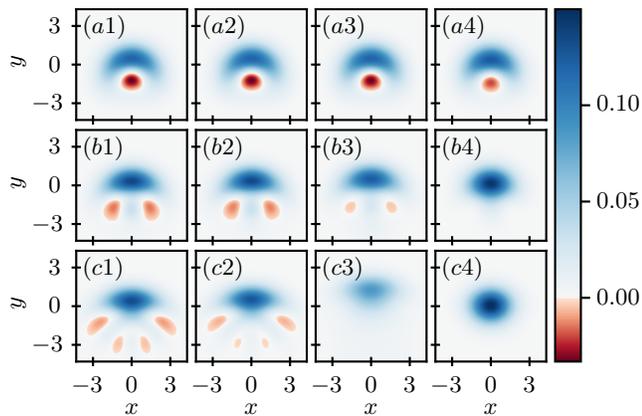
        \caption{Negative-valued Wigner distributions for temporal modes of (a) one, (b) two, and (c) four two-level atoms coupled to a cavity mode, with cavity decay rate $\kappa/g =2 $ and cavity driving strengths $\mathcal{E}/g=0.75N^{1/4}$. The spontaneous emission is increased from $(1)$ to $(4)$, with  $\gamma/g = \{2500^{-1},~1000^{-1},~250^{-1},~10^{-1}\}$ (with single-atom cooperativities $C=\{2500,~1000,~250,~10\}$), respectively.}
        \label{TLAGammawigs}
    \end{figure}

    Fig.~\ref{TLAGammawigs} presents Wigner distributions of the output field temporal modes for different numbers of two-level atoms coupled to the cavity, where the system is driven through the cavity mode. Coupling additional atoms to the cavity mode enables an increased number of correlated photon emissions into the output field. This manifests itself in the generation of Wigner-negative states exhibiting additional negative lobes in the Wigner distribution in direct proportion to the number of atoms. 
    
    So, clearly, these multi-atom systems allow for the realization of Wigner-negative states possessing more (negative) structure in their Wigner distributions. However, their steady state becomes progressively more sensitive to spontaneous emission with increasing $N$. This can be seen clearly in Fig.~\ref{TLAGammawigs}, where extremely strong coupling is required to retain appreciable negativity with an increasing number of atoms and an increasing level of spontaneous emission.
    
     The undesirable effect of independent-atom atomic spontaneous emission is related to its ability to transfer the atomic ensemble into states of non-maximal total spin. For example, with two atoms, spontaneous emission causes transitions from the maximally-coupled $S=1$ spin subspace to the $S=0$ singlet state, which is decoupled from the cavity mode and thus inhibits collective spin behavior. Additionally, the many-body control required to ensure identical atom-cavity coupling strengths is experimentally challenging.
     
     So, while it is clear from the presented results that collective-spin models are of interest for the generation of Wigner-negative light, steady-state implementation with an ensemble of atoms is problematic for practical cavity QED parameters. This prompts us to consider alternative implementations of collective-spin systems that are in some way more robust to, or minimize the effects of spontaneous emission.

\section{Collective Spin Systems: Idealized Models}
    Effective realizations of collective-spin models can, in principle, be implemented through Raman transitions within the hyperfine structure of a {\it single} atom, as proposed in \cite{Masson2017,Groiseau2021}. For example, as we shall describe below, suitably tailored, cavity-plus-laser assisted Raman transitions between magnetic sublevels of the $F=2$ hyperfine ground state of a ${}^{87}$Rb atom can achieve dynamics described by the Tavis-Cummings model for a collective spin-2 system coupled to the cavity mode. Importantly, with the fields far-detuned from the atomic excited states, the effects of atomic spontaneous emission are avoided, or at least made negligible in comparison to the desired dynamics.
    
    To be specific, here we consider a model of a single alkali atom coupled on the D1 line to a $\pi$-polarized cavity mode. The atom is also driven, in general on both the D1 and D2 lines, by suitably polarized laser fields. If the cavity and laser fields are all sufficiently far off-resonance from the transitions they couple to, then the atomic excited states are only weakly populated and can be adiabatically eliminated. Following the effective operator formalism of \cite{Reiter2012}, we are able to systematically derive a model for the resulting atomic ground-state dynamics. Details of the derivation are given in Appendix B.
    
    Of particular interest is the case where the atomic state is confined to a single hyperfine ground state of total angular momentum $F$, in which case the unitary evolution is governed by a Hamiltonian for a single spin-$F$ system. For the D1 line, coupled to a $\pi$-polarized cavity mode and a pair of oppositely circularly-polarized lasers, the Hamiltonian takes the general form
    \begin{equation}
        \begin{split}
        \hat{H}_{\rm D1} &= \omega\hat{a}^\dagger\hat{a} + \omega_0\hat{S}_z \\
        &~~~ + \lambda_-\left(\hat{S}_+\hat{a} + \hat{a}^\dagger\hat{S}_-\right) + \lambda_+\left(\hat{S}_-\hat{a} + \hat{a}^\dagger\hat{S}_+\right)\\
        &~~~ + \zeta_-\hat{Q}_{xz}\left(\hat{a}^\dagger + \hat{a}\right) + i\zeta_+ \hat{Q}_{yz}\left(\hat{a} - \hat{a}^\dagger\right) \\
        &~~~ + \tau\left(\hat{S}_+^2 +\hat{S}_-^2\right) +\left(\omega_q-\delta_q\hat{a}^\dagger\hat{a} \right)\hat{S}_z^2 ,
        \label{hamiltonianD1}
        \end{split}
    \end{equation}
    where $\hat{S}_i$ and $\hat Q_{ij}$ are spin-$F$ angular momentum and quadrupole operators, respectively, with $F$ determined by the particular hyperfine state that is populated. 
    
    Meanwhile, simultaneously illuminating the D2 line with lasers of all three polarizations gives a unitary evolution of ground states according to the general Hamiltonian
    \begin{equation}
        \begin{split}
        \hat{H}_{\rm D2} &=  \omega'_0\hat{S}_z +\omega'_q\hat{S}_z^2+ \zeta'_-\hat{Q}_{xz}+ i\zeta'_+ \hat{Q}_{yz}\\
        &~~~ + \lambda'\left(\hat{S}_+ + \hat{S}_-\right) + \tau'\left(\hat{S}_+^2 +\hat{S}_-^2\right).
        \end{split}
        \label{hamiltonianD2}
    \end{equation}
    
    All of the parameters appearing in these effective Hamiltonians can, in principle, be tuned by adjusting the frequencies and/or strengths of the laser and cavity fields. Explicit formulae for the parameters are given in Appendix B. Additionally, due to the large frequency separation between the D1 and D2 lines, the two Hamiltonians above can be implemented independently. This gives a versatile model for exploring a variety of effective, collective-spin dynamics within the internal states of a single atom. 
    
    \begin{figure}[htp]
        \centering
        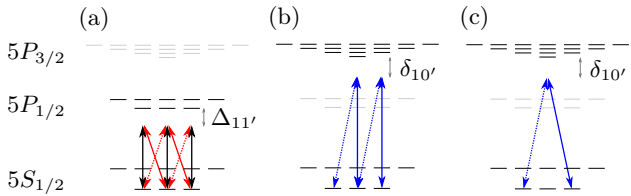
        \caption{Schematic for implementations of Raman transitions within the hyperfine structure of a ${}^{87}$Rb atom. Vertical arrows indicate $\pi$ transitions, and diagonal arrows represent $\sigma_+$ (dotted line) and $\sigma_-$ (solid line) transitions. (a) Cavity-assisted Raman transitions implemented on the D1 line, with a $\pi$-polarized cavity mode and $\sigma_\pm$-polarized auxiliary lasers. (b) One-step coherent Raman transitions driven on the D2 line, with a combination of $\pi$- and $\sigma_+$-polarized secondary lasers. (c) Two-step coherent Raman transitions driven on the D2 line through a pair of $\sigma_\pm$-polarized secondary lasers.}
        \label{RamanTransitions}
    \end{figure}

    The primary mechanism by which the cavity mode is populated with photons in our scheme involves a $\sigma_-$-polarized laser field exciting the D1 line and combining with the $\pi$-polarized cavity mode to drive Raman transitions towards lower $m_F$ states, as depicted in Fig.~\ref{RamanTransitions}(a). Engineering the Raman transition rate to be slower than cavity field decay ensures an essentially one-directional, cavity-mediated optical pumping process along the $m_F$ states. This also ensures the cavity output field will be representative of the specific $m_F$-increasing driving mechanism that we additionally implement via the other laser fields, in order to achieve a non-trivial steady state with a continuous cavity output field. In particular, we will consider three different driving mechanisms, each realising distinct effective models.
    
     Firstly, we consider driving the cavity mode with a laser field. This obviously increases the mean cavity photon number and therefore promotes cavity-assisted Raman transitions in both directions. This simply amounts to an effective Tavis-Cummings model with coherent cavity driving, as considered earlier, but now without spontaneous emission. We note at this point, however, that a nice feature of our effective-spin system is the alternative possibility of implementing direct driving of the collective-spin using standard, coherent Raman transitions, as illustrated in Fig.~\ref{RamanTransitions}(b). The states generated with such driving are essentially the same as for cavity driving though, so in this section we just consider cavity driving.
    
    Secondly, adding a $\sigma_+$-polarized auxiliary laser on the D1 line, shown also in Fig.~\ref{RamanTransitions}(a), gives a pathway for ``counter-rotating'', cavity-assisted Raman transitions, i.e., transitions that drive the atom towards {\em higher} $m_F$ states. This combination of driving lasers and cavity mode on the D1 line allows the realization of an effective Hamiltonian corresponding to the celebrated Dicke model.
    
    Finally, a pair of oppositely circularly-polarized lasers on the D2 line facilitate coherent Raman transitions that raise and lower the atomic state by $\Delta m_F=\pm2$. This yields a ``two-step'' atomic driving model, which is, notably, not readily attainable with ensembles of spin-1/2 atoms.

    In the following three subsections, we consider these three different forms of driving in our single-atom scheme in the limit of very large detunings of the fields acting on the D1 line; in particular, detunings that far exceed the excited-state hyperfine splitting on the D1 line. In this regime, the terms in the last two lines of the Hamiltonian (\ref{hamiltonianD1}) become negligible, and the remaining atom-cavity interaction terms in the second line are simply of the form of the Tavis-Cummings and anti-Tavis-Cummings models, respectively. Hence, we are able to realize simple, idealized models of collective spin systems coupled to a cavity field mode.

    In practice, this regime of very large detunings also demands the limit of extremely large, resonant interaction strengths between the atom and cavity mode \cite{Groiseau2021}. The viability of these models under the constraints of modern, optical cavity QED experimental parameters will be explored in Section V of this paper, where a full, single-alkali-atom cavity QED model of the scheme is investigated.

\subsection{Tavis-Cummings Model with Cavity Driving}
\begin{figure}[htp]
        \centering
        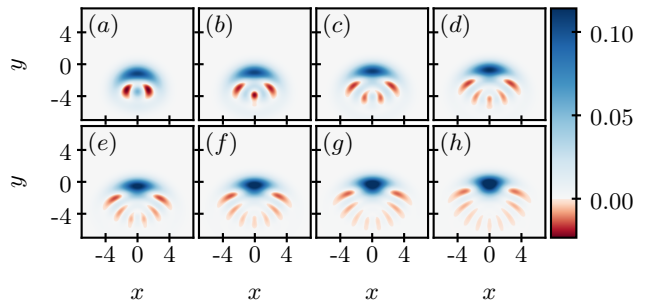
        \caption{Wigner distributions for the coherently-driven Tavis-Cummings model with different collective-spin lengths. The effective spin increases (a to h) in steps of 1/2, from $F = 1$ to $F=9/2$, with $\kappa=2(2F)^{1/2}\lambda_-$, and $\mathcal{E} = 0.75 (2F)^{3/4}\lambda_-$.}
        \label{TCMWigner}
    \end{figure}

    An effective Tavis-Cummings model with coherent cavity driving takes the form 
    \begin{equation}\label{eq:dTCM}
        \hat{H}_{TC} = \lambda_-\left(\hat{S}_+\hat{a} + \hat{a}^\dagger\hat{S}_-\right) + \mathcal{E}\left(\hat{a}+ \hat{a}^\dagger\right).
    \end{equation}
    As with the atomic ensemble realisation, the superposition states generated from this system depend on the total effective spin length. This is shown in Fig.~\ref{TCMWigner}, where, as the effective spin-length increases, additional negative lobes appear in the Wigner distribution. Without spontaneous emission present, this growing structure remains clearly evident. 
    
    Furthermore, the increasing number of lobes compensates for their diminishing depth. The total negative volume for the ideal two-level atom in Fig.~\ref{QBCOPBCompare}(a) is $0.041$. The Wigner distributions in Fig.~\ref{TCMWigner} have similar negative volumes; in fact, all are $\mathcal{N}=0.04$ to one significant figure. More precisely, the negative volumes in Figs.~\ref{TCMWigner}(a, c , g) are $\mathcal{N} = \{0.040,~0.043,~0.040\}$, respectively. 
    
    The volumes are not directly comparable as the cavity driving strength is increased in conjunction with the spin-length, to ensure the systems are sufficiently well excited. Nonetheless, the sustained negativity shows that the Tavis-Cummings model could be suitable as a bright source of Wigner-negative light. 
    
\subsection{Dicke Model}
\begin{figure}[t]
        \centering
        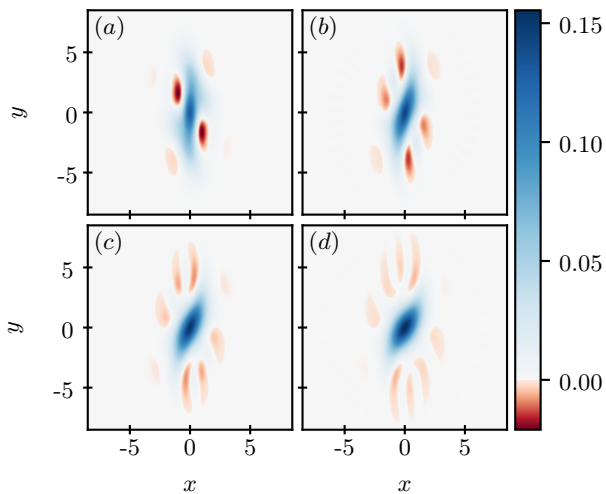
        \caption{Wigner distributions of temporal modes from an imbalanced Dicke model with $\lambda_+ = 0.5\lambda_-$. The effective spin length is increased (a to d) in steps of 1, from $F = 1$ to $F=4$, and  with parameters of $\kappa=2(2F)^{1/2}\lambda_-$, and $\omega_0 = 0.4 (2F)^{-1/2}\lambda_-$.}
        \label{IdealDickeWig}
    \end{figure}

    The Dicke model is another example of a well-studied collective-spin system. For the generation of Wigner-negative light, it is desirable to realize an ``unbalanced'' Dicke model, such that $\lambda_+\neq\lambda_-$. The Dicke Hamiltonian we consider is of the form 
    \begin{equation}
        \hat{H}_D = \omega_0\hat{S}_z +\lambda_-\left(\hat{S}_+\hat{a} + \hat{a}^\dagger\hat{S}_-\right) + \lambda_+\left(\hat{a}\hat{S}_- + \hat{S}_+\hat{a}^\dagger\right).
        \label{DickeHam}
    \end{equation}
    In the particular regime where $\kappa>\lambda_->\lambda_+$, the system tends to emit photons in pairs. The atomic state is raised by a $\lambda_+$ transition, then lowered by a $\lambda_-$ transition. Each of these transitions generates a cavity photon, which decays into the output field before re-absorption by the atom is possible, so a pair of photons is emitted each time a cycle is completed. This leads to temporal mode states that favor superpositions of even-numbered photon states, resulting in Wigner distributions as shown in Fig.~\ref{IdealDickeWig}. 
    As the effective spin length is increased, there are further, simultaneous iterations of the cycle, which adds larger, even numbers of photons to the field. 
    
\subsection{Two-Step Collective Atomic Driving}
    As mentioned earlier, a nice feature of the single-atom implementation of effective-spin systems is the availability of coherent two-step Raman transitions, which change the atomic state by $\Delta m_F = \pm2$. Driving these coherent Raman transitions on the D2 line, to accompany the cavity-assisted Raman transitions on the D1 line, generates a Hamiltonian with quadratic driving of the collective atomic spin,
    \begin{equation}
        \hat{H}_{TS} = \lambda_-\left(\hat{S}_+\hat{a} + \hat{a}^\dagger \hat{S}_- \right) + \tau\left(\hat{S}_+^2 + \hat{S}_-^2\right).
    \end{equation}
    In a limit where $\lambda\gg\Omega$, the atom moves through a cycle of transitions, excited first by the two-step driving, before decay through consecutive cavity-assisted Raman transitions. Each iteration of this cycle deposits two photons in the cavity field, which ultimately escape into the cavity output field.
    
\begin{figure}[htp]
        \centering
        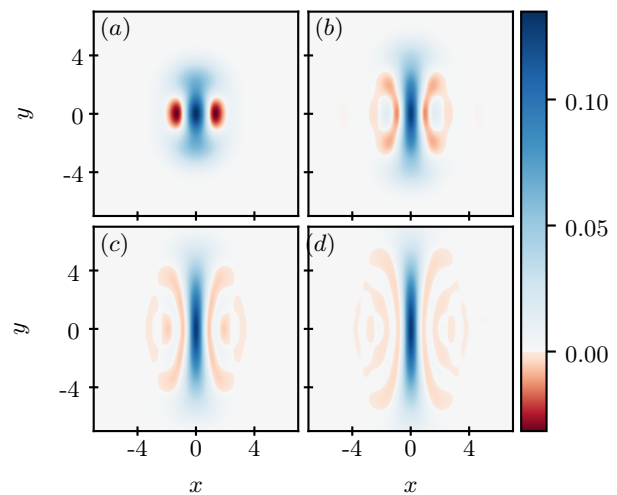
        \caption{Wigner distributions for the Tavis-Cummings model with two-step collective atomic driving, with the effective spin length increased (a to d) in steps of 1, from $F = 1$ to $F=4$, and with parameters of $\kappa=2(2F)^{1/2}\lambda_-$, and $\tau = 0.15(2F)^{-1/2}\lambda_-$.}
        \label{IdealTPDWig}
    \end{figure}

    In the same parameter regime, larger superpositions of even-numbered photon states can be achieved by increasing the atomic spin length. Each additional atomic state that is added opens the possibility for further iterations of the cycle of transitions, leading to a state of the output temporal mode that is loosely approximated by a superposition of even-numbered photon states up to $2(2F-1)$. Fig.~\ref{IdealTPDWig} shows Wigner distributions that are generated as the spin length is increased from $F=1$ to $F=4$.
     
    While it is clear that the extent of Wigner negativity decreases quickly with increasing $F$, it is also evident that, in this same limit, the system produces significant quadrature squeezing of the temporal output field mode. In fact, the uncertainty $\Delta\hat X_f$ is reduced below the vacuum state level of 1 to the value 0.46 for $F=3$ and 0.30 for $F=4$. 



\section{Collective Spin Systems: A Single Alkali Atom}
    
    The idealized models show promise for the generation of Wigner-negative states in the steady-state output from driven cavity QED systems. In practice, however, some level of imperfection is expected in any cavity QED experiments based on actual atoms. The ramifications can be especially important to steady-state behavior; even very weak processes can strongly impact the state after a sufficiently long time. In this section, the viability of implementing the models is explored with realistic cavity QED parameters, employing a single ${}^{87}$Rb atom. A model for the D1 and D2 lines of the ${}^{87}$Rb atom is written in a basis of the hyperfine Zeeman sublevels. By carefully selecting the operating regime, this is used to replicate the dynamics produced in the idealized models, both for the case of a single two-level (spin-1/2) atom and for a larger-spin, collective atomic system ($F=1$ or $F=2$ for ${}^{87}$Rb). The same effects can of course be explored in other alkali atoms; for example ${}^{133}$Cs, which offers $F=3$ and $F=4$ dynamics in appropriate operating regimes.

    For the purpose of consistency, we will explore the implementation of the effective models based on Raman transitions in a strong coupling regime, with cavity QED parameters $\{g,~\kappa,~\gamma_{D1},~\gamma_{D2}\}=\{500,~50,~5.750,~6.066\}~2\pi\text{MHz}$, unless specified otherwise. Our choice of parameters is encouraged by recent experiments with micro- and nano-scale optical cavities \cite{Gehr2010,Samutpraphoot2020}, and corresponds to a large single-atom cooperativity of $C=1739$. As we shall demonstrate, tuning the driving configurations provides sufficient flexibility to achieve desirable operating regimes for each of the models we consider.
    
\subsection{Atomic Model}
    A ${}^{87}$Rb atom has two hyperfine ground electronic states ($F=1,2$), each with $2F+1$ Zeeman sublevels, which we label by $m_F$. The D1 line involves transitions to two hyperfine excited states ($F=1', 2'$), while the D2 line has four ($F=0', 1', 2', 3'$). Modeling of transitions driven by an external field can be written in a complete basis of the Zeeman sublevels. A compact notation is provided by the dipole transition operators,
    \begin{equation}
        \hat{D}_q^{J'}(F,F') =\sum_{m_F =-F}^F C^{J',F,F'}_{m_F,q}\ket{F,m_F}\bra{J',F',m_F+q}
    \end{equation}
    where $q\in\{0,~\pm1\}$ reflects the transfer of $z$-projected angular momentum, and $C^{J',F,F'}_{m_F,q}$ are Clebsch-Gordan coefficients, normalised to 1 for the D2 cycling transiton. 
    
    The fine-structure splitting is sufficiently large that the two D lines can be treated independently. The D1 line is driven with circularly-polarized lasers and coupled to a linearly-polarized cavity mode, as described by the Hamiltonian 
    \begin{equation}
        \hat{H}_{D1} =\sum_{F,F'}\left( \frac{\Omega_+}{2}\hat{D}^{1/2}_{+} +g\hat{a}^\dagger\hat{D}^{1/2}_{0} + \frac{\Omega_-}{2}\hat{D}^{1/2}_{-} + {\rm H.c.} \right) .
    \end{equation}
    
    Additionally, D2-line transitions are driven with a combination of polarized lasers to facilitate coherent Raman transitions between ground-state Zeeman sublevels within a single hyperfine state. A general Hamiltonian describing this driving can be written as
    \begin{equation}
        \hat{H}_{D2} =\sum_{F,F'}\left( \frac{\Omega^{D2}_+}{2}\hat{D}^{3/2}_{+} +\frac{\Omega^{D2}_0}{2}\hat{D}^{3/2}_{0} + \frac{\Omega^{D2}_-}{2}\hat{D}^{3/2}_{-} + {\rm H.c.} \right).
    \end{equation}
    A Hamiltonian for the total atom-cavity system can then be written as 
    \begin{equation}
        \hat{H} =\hat{H}_0+\hat{H}_{D1}+\hat{H}_{D2},
    \label{FullHamiltonian}
    \end{equation}
    where $\hat{H}_0$ contains the bare atomic and cavity energies, which can also include Zeeman shifts resulting from an external magnetic field. Atomic spontaneous emission is considered to be independent for each D line and each polarization, with collapse (emission) operators of the form 
    \begin{equation}
        \hat{L}^{J'}_q = \sqrt{\gamma_{J'}}\sum_{F,F'}\hat{D}^{J'}_{q}(F,F').
    \end{equation}

Finally, we point out explicitly that, while each laser or cavity field couples to only one of the D lines, within that line they are assumed to couple to all allowed transitions (i.e., to all allowed $F\leftrightarrow F'$ transitions within the D line). In the schematics of the various configurations that follow, we draw only the dominant atom-field couplings.

\subsection{Two-Level Atom Regime}
Two-level behavior can be realized within the D2 line of ${}^{87}$Rb by confining the population to the cycling transition. For the cavity this ideally requires a circularly-polarized mode and zero birefringence, however this is particularly difficult to achieve in the micro- or nano-cavity systems that we have in mind and that realize the requisite large coupling strengths. The natural eigenmodes that characterize such cavities are linearly polarized, which can still couple to the cycling transition, but also allow transitions out of the closed cycle, thus inhibiting idealized two-level behavior. 
To this end, in this section we consider a more moderate coupling strength of $g=150\tpmhz$, to allow for more efficient confinement to the chosen levels. Since we assume the cavity mode to be horizontally-polarized, coupling to the cycling transition is actually reduced to $g/\sqrt{2}$.

\begin{figure}[htp]
        \centering
        \input{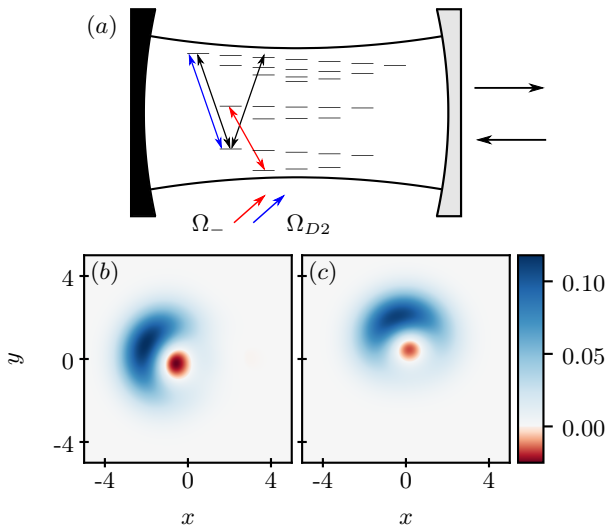}
        \caption{(a) Schematic and (b, c) Wigner distributions, for realising a two-level atom in the hyperfine structure of the $^{87}$Rb D2 line, with $\{g,~\gamma_{D2}\}=\{150,~6.066\}$. (b) In the quasi-bad-cavity regime, with $\{\kappa,~\Omega_{D2}\}=\{200,~60\}\tpmhz$ ($C = 37$), and the driving laser resonant with the cycling transition. (c) In the photon-blockade regime, with $\{\kappa,~\Omega_{D2}\}=\{40,~40\}\tpmhz$ ($C=185$), and the laser tuned above the cycling transition resonance by $-g/\sqrt{2}$. Isolation of the two levels is achieved by applying a $-150$~G magnetic field to generate Zeeman shifts in the energy levels, and a $\sigma_-$ repumping beam with $\Omega_- = g/2$.}
        \label{Rb87TLA}
    \end{figure}

A magnetic field that induces suitably large Zeeman shifts is also used to further suppress transitions outside of the desired states, while a repumping laser acting on the D1 line is required to maintain the system in a two-level regime in steady state.
The schematic in Fig.~\ref{Rb87TLA}(a) illustrates the repumping and Zeeman shifts which conspire to keep the atom confined to the cycling transition. 
Output temporal mode Wigner distributions are shown in Figs.~\ref{Rb87TLA}(b, c) for representative sets of parameters that enable good confinement of the atomic state to the cycling transition and yield Wigner-negativity in both the quasi-bad-cavity and photon-blockade regimes, respectively, as predicted earlier for the idealized Jaynes-Cummings model. A confinement of $95.7\%$ is achieved in Fig.~\ref{Rb87TLA}(b), while $90.0\%$ of the population occupies the cycling transition in Fig.~\ref{Rb87TLA}(c).

We note that these specific results employ a reasonably strong magnetic field of $B=-150$~G. 
Due to the large magnetic field strengths being employed, it is necessary to include the full Zeeman effect, beyond the linear approximation, by expanding 
\begin{equation}
    \hat{H}_Z = -\mu\cdot B\approx\mu_B (J+S)\cdot B.
\end{equation}
While the full Hamiltonian is used for simulations, the probe and repumping laser frequencies are chosen by using the linear Zeeman approximation $\Delta\omega =  g_F m_F B \mu_B$. 

\begin{figure}[htp]
        \centering
        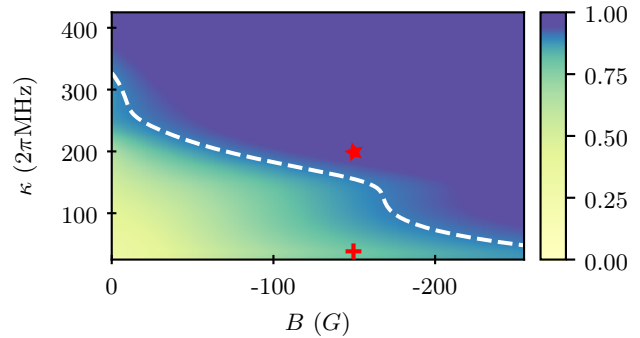
        \caption{Total steady-state population confinement to the cycling transition in a ${}^{87}$Rb atom coupled ($g=150\tpmhz$) to a horizontally-polarized cavity mode and driven by a $\sigma_-$ laser, tuned with the linear Zeeman shift to the upper photon-blockade resonance. The dashed white line shows 90\% confinement to the cycling transition. The red star and cross show parameters chosen for the Wigner distributions in Figs.~\ref{Rb87TLA}(b) and (c), respectively.}
        \label{TLAConfinement}
    \end{figure}

Fig.~\ref{TLAConfinement} shows how well the atom is confined to the cycling transition, for different cavity field decay rates and magnetic field strengths. Confining the atom to the most negative $m_F$ ground state works best with a negative-valued magnetic field, such that the transition shifts away from other resonances. For the same reason, the blockade effect is best isolated by tuning to the higher-frequency single-photon resonance. 

For larger field decay rates, the system can give high-quality two-level behavior with much weaker Zeeman shifts. Isolating the photon blockade regime requires a narrower cavity linewidth, and so a strong magnetic field is needed to confine the population. In principle, ${}^{133}$Cs is also a nice candidate for realizing the two-level behavior with a linearly-polarized cavity mode, due to a more favorable ratio of Clebsch-Gordan coefficients into, compared to out of, the cycling transition.

\subsection{Effective-Spin Models}
    In an appropriate driving regime, where the excited states are only weakly populated, an effective evolution of the ground state populations is derived through adiabatic elimination of the excited states. The resulting Hamiltonian takes the form of effective spin interactions within each hyperfine ground level given by Hamiltonians (\ref{hamiltonianD1}) and (\ref{hamiltonianD2}) for D1 and D2 line transitions, along with terms that couple (off-resonantly, and thereby weakly) the two ground state manifolds. 
    
\begin{figure}[htp]
        \centering
        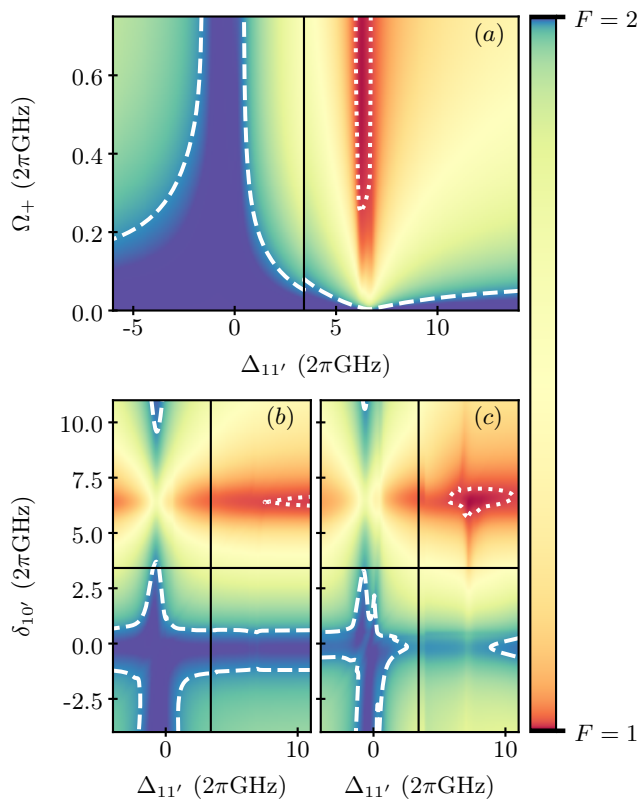
        \caption{Steady-state distribution of atomic population amongst the two ground-state hyperfine manifolds. White dashed (dotted) contours indicate $99\%$ confinement to the $F=2$ ($F=1$) state. In all cases, the D1 line is driven with $\Omega_- = 1~2\pi\text{GHz}$, along with coupling to the $\pi$-polarized cavity ($\{g,~\kappa\}=\{500,~50\}\tpmhz$). This is augmented with (a) driving the D1 line with an $\Omega_+$ laser, or (b, c) driving the D2 line with a pair of lasers with Rabi frequency $\Omega_{D2}=1~2\pi\text{GHz}$, where the two beams are (b) $\sigma_\pm$-polarized, or (c) $\sigma_+$- and $\pi$-polarized. Black lines indicate changing the magnetic field depending on the target state, according to the linear Stark shifts in Appendix B. In all cases, only moderate magnetic fields are required, such that $|B|\lesssim 20$~G.}
        \label{AllScansUnits}
    \end{figure}

    In general, the steady-state atomic population can be arbitrarily distributed amongst both ground state manifolds. In order to realize effective-spin dynamics, it is necessary to identify operating regimes where the population is confined to a single hyperfine level. The implementation of cavity coupling in the models considered here will cause the atom to favor the (upper) $F=2$ state, due to accumulation of population in the maximal-projection state ($m_F=\pm2$). If the driving mechanism increasing $m_F$ is sufficiently strong, then the (lower) $F=1$ state can be populated by tuning the lasers to act as selective repumping beams. The far-off resonant nature of the driving is particularly convenient in this context, as the detunings can be chosen such that the lasers can both implement effective dynamics in the target state, while providing a repumping mechanism out of the undesired state. Fig. \ref{AllScansUnits} shows the steady-state atomic populations, calculated from the full atomic model including the excited states, for the driving setups to be considered. 
    
    In all cases, the steady-state population can be very well confined (>99\%) to the chosen hyperfine level. As shown in Fig.~\ref{AllScansUnits}, confining the atom to the $F=2$ level means tuning the lasers such that $\{\Delta_{11'},~\delta_{10'}\}\approx0$, so that transitions out of the $F=1$ state are close to resonant. Conversely, tuning the lasers such that $\{\Delta_{11'},~\delta_{10'}\}\approx\omega_g$, where $\omega_g\approx6.8\tpmhz$ is the ground state hyperfine splitting of ${}^{87}$Rb, will cause transitions out of the $F=2$ state to be near-resonant, confining the population to $F=1$.
    
    The presence of light fields shifts the energies of the atomic states, the dominant effect being a linear Stark shift. The shifts can be cancelled by applying a suitable external magnetic field, which induces a linear Zeeman shift. The shifts are different for the two ground-state hyperfine levels, so when confining the atom to a single hyperfine manifold, we choose the magnetic field such that the linear shifts in the target level will approximately cancel. The magnetic field strength is chosen according to the linear shift term in the effective Hamiltonians (\ref{hamiltonianD1}) and (\ref{hamiltonianD2}).

    For consistency between effective models, a common set of parameters is picked for implementation of the cavity coupling in a given $F$ level. Fields on the D1 line are always tuned such that $\Delta_{11'} = 6200\tpmhz$ and $\Delta_{11'} = -750\tpmhz$ for the $F=1$ and $F=2$ models respectively, and the $\sigma_-$-polarized laser is set to a strength of $\Omega_- = 1000\tpmhz$. The models which include a pair of lasers on the D2 line are tuned such that $\delta_{10'} = 6600\tpmhz$ and $\delta_{10'} = -150\tpmhz$ for the $F=1$ and $F=2$ models respectively. Additionally, the D2 laser pair are also chosen to have the same Rabi frequency, $\Omega_{D2}$. This means that different models can be implemented by interchanging only the targeted D line or the polarisation and Rabi frequency of the $m_F$-increasing driving lasers.

\subsubsection{Driven Tavis-Cummings Model}

\begin{figure}[t]
        \centering
        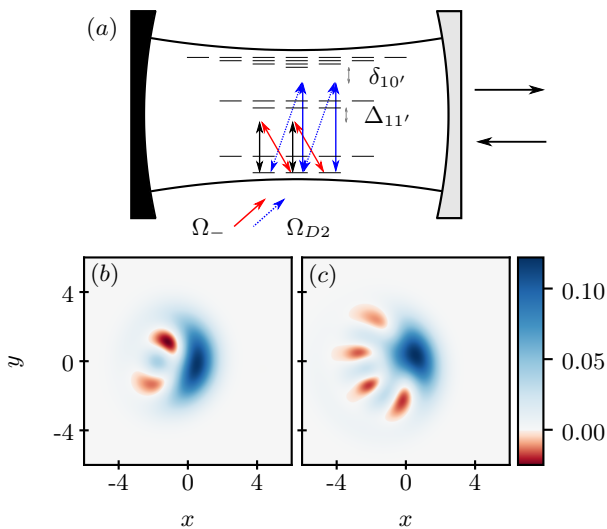
        \caption{(a) Schematic for implementation of a Tavis-Cummings model, with cavity coupling ($\{g,~\kappa\}=\{500,~50\}\tpmhz$) on the D1 line, and collective driving on the D2 line. Wigner distributions in the TCM regime are shown for (b) the $F=1$ model, with $\{\Omega_{D2},~\Delta_{11'},~\delta_{10'}\}=\{270,~6200,~6600\}\,2\pi\text{MHz}$ and $B = 8.9$~G, and (c) an $F=2$ model, with $\{\Omega_{D2},~\Delta_{11'},~\delta_{10'}\}=\{360,~-750,~-150\}\,2\pi\text{MHz}$, and $B = -9.26$~G.}
        \label{fig:enter-label}
    \end{figure}

    The cavity driving employed in the idealized model of (\ref{eq:dTCM}) becomes challenging to simulate in the full-atomic-structure model, owing to the large basis size required to adequately model the cavity mode in numerical calculations. So, instead, a one-step coherent Raman transition is used to implement collective driving, as shown in Fig.~\ref{RamanTransitions}(b). As with the previous models, results are equivalent regardless of the chosen driving mechanism. The full excitation scheme therefore takes the form depicted in Fig.~\ref{fig:enter-label}(a), for realization of the effective spin-1 model.

    As demonstrated in Figs.~\ref{fig:enter-label}(b, c), the Wigner distributions follow precisely (apart from a rotation in phase space) the patterns predicted from the idealized model for spin-1 and spin-2 collective systems, where the number of negative lobes corresponds directly to the effective spin length.

\subsubsection{Dicke Model}
\begin{figure}[t]
        \centering
        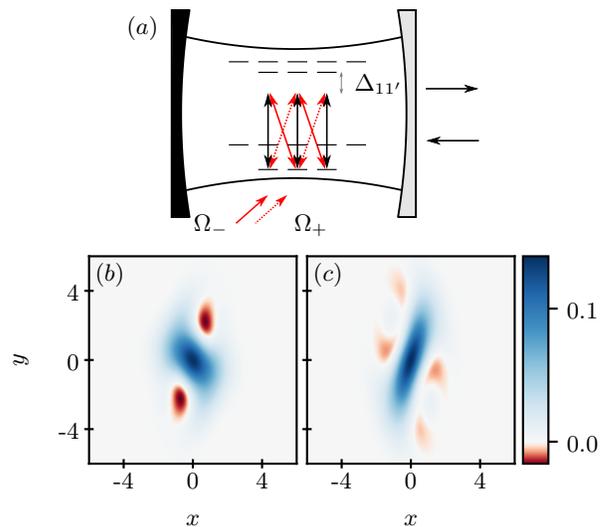
        \caption{(a) Implementation of the Dicke model in a ($\{g,~\kappa\}=\{500,~50\}\tpmhz$) single-atom cavity QED system, with counter- and co-rotating terms realized on the D1 line. Wigner distributions are shown in $\{(b),~(c)\}$ for $\Omega_+=0.35\Omega_-$ and $\Delta_{11'} = \{6200,~-750\}~2\pi\text{MHz}$, $B = \{7.2,~-7.3\}$~G, corresponding to the $F=1$ and $F=2$ models, respectively.}
        \label{FullDickeWignerSchematic}
    \end{figure}

    Augmenting the D1 line excitation with a $\sigma_+$-polarized laser provides a pathway for $m_F$-increasing cavity-assisted Raman transitions, as illustrated in Fig.~\ref{FullDickeWignerSchematic} (a). This gives a counter-rotating term in addition to the co-rotating transition term, both mediated by the cavity mode, producing unitary dynamics equivalent to the Dicke model.

    For the purposes of this work, as discussed earlier, the model is chosen to be unbalanced, i.e., $\lambda_->\lambda_+$. As with all the models, if the driving is too asymmetric, the atom is optically-pumped to the $m_F=\pm2$ state. This can be seen from Fig.~\ref{AllScansUnits}(a), where a sufficiently strong $\Omega_+$ is required to confine the population to the $F=1$ state. For the parameters we explore, the imbalance is acceptably small that the population confinement can be readily achieved. In principle, an explicit repumping laser could be added to the model (for example on the D2 line with $\delta_{10'}\approx\omega_g$) to allow a more imbalanced model in the $F=1$ state, if desired. 
    
    The term in Hamiltonian (\ref{DickeHam}) proportional to $\omega_0$ can be controlled by tuning the external magnetic field. With all of the parameters specified appropriately, the predicted states, shown in Fig.~\ref{FullDickeWignerSchematic}(b, c), closely reproduce  those found with the relevant idealized model.

\subsubsection{Two-Step Collective Atomic Driving}
    A pair of circularly-polarized lasers on the D2 line will drive coherent Raman transitions between ground states with $\Delta m_F = \pm2$. Accompanied with the D1 cavity-assisted Raman transitions, the two-step nature of driving manifests itself in the tendency for photons to be emitted from the cavity in pairs. 
    
    The excited-state pathways for two-step transitions actually interfere destructively when the lasers are tuned very far from resonance. This can be seen in the explicit formulae for the effective parameters given in Appendix B; when the detunings are much larger than the excited-state hyperfine splittings, the $\mathcal{T}$ term of Eq.~(\ref{GeneralHam}) vanishes. This suggests that the two-step driving should be implemented closer to resonance, however, this is not suitable for the $F=1$ model, due to the repumping requirements. So, the two-step driving is still implemented with far-off-resonant lasers, but we increase the laser strengths to compensate and give an adequate rate for this process.
    
    Due to the tendency of the D1 optical pumping mechanism to favor the (upper) $F=2$ state, the spin-2 results can also be achieved when the $D2$ line lasers are much closer to ($F=2$) resonance, i.e., $\delta_{10'}\approx \omega_g$. This allows for an equivalent model to be implemented in practice with significantly weaker lasers on the D2 line. 

\begin{figure}[t]
        \centering
        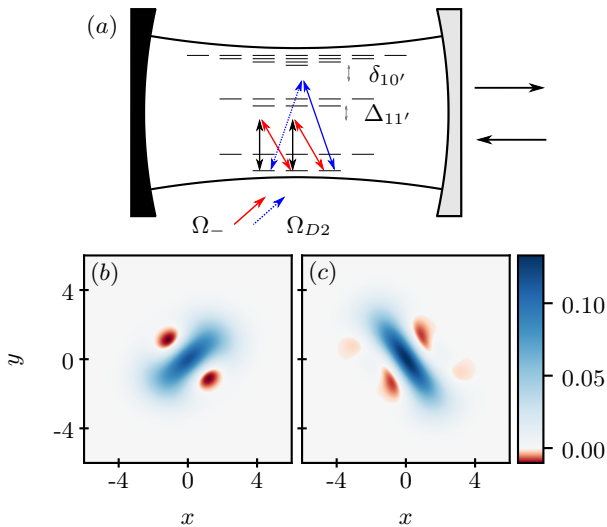
        \caption{(a) Schematic for the implementation of two-state driving on the D2 line, with ($\{g,~\kappa\}=\{500,~50\}\tpmhz$) cavity-assisted transitions on the D1 line. Wigner distributions from the two-Photon Driving model are shown for (b) the $F=1$ model, with $\{\Omega_{D2},~\Delta_{11'},~\delta_{10'}\}=\{1750,~6200,~6600\}\,2\pi\text{MHz}$ and $B = 8.2$~G, and (c) an $F=2$ model, with $\{\Omega_{D2},~\Delta_{11'},~\delta_{10'}\}=\{1550,~-750,~-150\}\,2\pi\text{MHz}$ and $B = -8.1$~G.}
        \label{FullTPDWignerSchematic}
    \end{figure}

    The implementation of two-step driving also introduces an appreciable quadratic Stark shift, which cannot generally be cancelled with an external magnetic field. Fortunately, the magnitude of shift is relatively small, as long as the two-step driving is weak. In principle, one can also find operating regimes where the D1 quadratic Stark shift is engineered to cancel the equivalent D2 shifts. For the results presented here, this is not implemented, as clearly the states in Fig.~\ref{FullTPDWignerSchematic}(b,c) are already very similar to those predicted by the idealized model.

    As in the simpler models, the Wigner distributions take the form of even-numbered photon superpositions, and demonstrate an increased squeezing of the central lobe for longer spin-length. The squeezing axis is rotated, but the reduced variance can be quantified through the generalised quadrature operator,
    \begin{equation}
        \hat{X}_\theta = \hat{a} e^{-i\theta} + \hat{a}^\dagger e^{i\theta},
    \end{equation}
    for a suitable phase angle $\theta$.
    
    For the $F=1$ model state, Fig. \ref{FullTPDWignerSchematic}(b), a maximally-squeezed variance of $0.9218$ is attained, while the $F=2$ state in Fig. \ref{FullTPDWignerSchematic}(c) exhibits a minimal variance of $0.5128$. Note that these values differ from the ideal model numbers given earlier, since the models are considered under different operating parameters. The two-state driving model is a promising avenue for a single-atom source of strongly-squeezed light. This will be explored further for ${}^{87}$Rb, as well as other alkali atoms, in a future work.
    
\section{Conclusion}
    In this work we have examined the open, driven Jaynes-Cummings model and some of its multi-atom variants from a new perspective, that of Wigner-distribution negativity in states of temporal modes of the cavity output field. Remarkably, these models once again open new and intriguing results, which we hope may pique the interest of theorists and experimentalists alike. Clearly, there is scope for further, more detailed theoretical investigation of the origins of the negative structure seen in the various Wigner distributions presented in this work. Meanwhile, our use of experimentally relevant parameters suggests that our schemes may be viable practical sources of Wigner-negative light using single-atom, optical cavity QED setups. The models and results presented are, of course, also of potential interest to the burgeoning field of circuit QED, from which, indeed, we gained much of our original motivation for pursuing this work. 

\section*{Acknowledgments}
    The authors wish to acknowledge the use of New Zealand eScience Infrastructure (NeSI) high performance computing facilities, consulting support and/or training services as part of this research. New Zealand's national facilities are provided by NeSI and funded jointly by NeSI's collaborator institutions and through the Ministry of Business, Innovation and Employment's Research Infrastructure programme (URL https://www.nesi.org.nz). The authors also acknowledge support
    from the Marsden Fund of the Royal Society of New Zealand (MFP-21-UOA-280). A. E. thanks Blair Blakie and the Department of Physics at the University of Otago, for their hospitality and support during the preparation and write-up of this manuscript.

\section*{Disclosures}
    The authors declare no conflicts of interest.

\bibliography{References}

\begin{thebibliography}{47}
\bibitem [{mee()}]{mee}%
  \BibitemOpen
  \href@noop {} {}\bibinfo {note} {Scripts can be made available upon
  reasonable request to the authors.}\BibitemShut {Stop}%
\end{thebibliography}

\appendix

\section{Capture Cavity Method}
The capture cavity method is based on the input-output theory for quantum pulses \cite{Kiilerich2019}. It describes a gedankenexperiment, where the output from a quantum system is coupled into a ``capture'' cavity, with a time-dependent coupling strength. The cavity will asymptotically acquire the state of a temporal mode with an envelope $f(t)$, provided the coupling is given by 
\begin{equation}
    \kappa_f(t) = -\frac{f^*(t)}{\sqrt{\int_0^t|f(t')|^2dt'}}.
\end{equation}
The required coupling has a singularity at $t=0$, so in practice we choose some small $\epsilon>0$ to write
\begin{equation}
    \kappa_f(0) = -\frac{f^*(0)}{\sqrt{\int_0^\epsilon|f(t')|^2dt'}}.
\end{equation}
Defining $\hat{b}$ ($\hat{b}^\dagger$) as the annihilation (creation) operator for the capture-cavity mode, the system can then be modelled as a cascaded quantum system, with a Hamiltonian
\begin{equation}
    \hat{H}_{\text{cap}} = \hat{H} +\frac{i}{2}\left(\sqrt{2\kappa\kappa_f(t)}\, \hat b^\dag \hat a - {\rm H.c.}\right),
\end{equation}
and modified collapse operator
\begin{equation}
    \hat{L}_f(t) = \sqrt{2\kappa}\,\hat{a} + \kappa_f^*(t)\hat{b}(t).
\end{equation}
The final temporal mode state is calculated numerically, by integrating the time-dependent master equation over the temporal mode envelope function. Partially tracing over the source system gives a reduced density operator for the capture cavity, equivalent to the desired temporal mode state.

\section{Adiabatic Elimination}
Applying the effective operator formalism to adiabatic elimination of the excited states gives analytic expressions for effective ground-state dynamics \cite{Reiter2012}. A brief outline of the method applied to the ${}^{87}$Rb D lines is given here. Treating the fine structure doublet independently means the process can be applied separately to each of the D lines. For each D line, the atomic Hamiltonian \ref{FullHamiltonian} is separated into terms which couple within the ground states ($\hat{H}_\text{g}$) and excited states ($\hat{H}_\text{e}$), along with terms which couple between them ($\hat{V}_\pm$), 
\begin{equation}
    \hat{H} = \hat{H}_\text{e} + \hat{H}_\text{g} + \hat{V}_+ + \hat{V}_-.
\end{equation}
For a ${}^{87}$Rb atom, the ground state Hamiltonian is 
\begin{equation}
    \hat{H}_g =\hat{H}_\text{Zg}+ \sum_{F,m_F}\omega_F\ket{m_F,F}\bra{m_F,F},
\end{equation}
where $\hat{H}_\text{Zg}$ describes Zeeman effect on the ground levels, and $\omega_F = 0$ ($\omega_g$) for the $F=1$ ($F=2$) state. The excited state Hamiltonian is 
\begin{equation}
        \hat{H}_e =\hat{H}_\text{Ze}+ \sum_{F',m_F}\zeta_{1F'}\ket{F',m_F}\bra{F',m_F},
    \end{equation}
where $\hat{H}_\text{Ze}$ is the excited-state Zeeman effect, and $\zeta_{1F'} = \{\Delta_{1F'},~\delta_{1F'}\}$ represents the relevant detunings. The detunings are labelled $\Delta_{FF'}$ with $F'=\{1,~2\}$, and $\delta_{1F'}$, with $F'=\{0,~1,~2,~3\}$, for D1 and D2 states respectively, defined as the coupling field detuning from the $F\rightarrow F'$ transition. 
The formalism gives an effective master equation for evolution of the ground states \cite{Reiter2012}
\begin{equation}
    \dot{\hat{\rho}} = -i[\hat{H}_{\text{eff}},\hat{\rho}] + \mathcal{D}[\sqrt{2\kappa}\hat{a}]\hat{\rho}+ \sum_{q} \mathcal{D}[\hat{L}_{\text{eff}}^q]\hat{\rho},
\end{equation}
Contributions from independent fields (i.e each polarization) are delineated by $q$, and excitation from different ground states labelled $F$. Each D line can couple to three distinct polarizations ($\pi$ and $\sigma_\pm$), from each of two hyperfine ground levels. The effective operators are
\begin{equation}
\begin{split}
    \hat{H}_\text{eff} &=\hat{H}_g -\frac{1}{2}\left[\hat{V}_-\sum_{F,q} \left(\hat{H}_\text{NH}^{(F,q)}\right)^{-1}\hat{V}_+^{(F,q)} + \text{H.c}\right],\\
    \hat{L}^q_\text{eff} &= \hat{L}_q\sum_{F,q'} \left(\hat{H}_\text{NH}^{(F,q')}\right)^{-1}\hat{V}_+^{(F,q')},
\end{split}
\end{equation}
with
\begin{equation}
    \left(\hat{H}_\text{NH}^{(F,q)}\right)^{-1} \equiv \left(\hat{H}_\text{NH}-\omega_F -\omega_q\right)^{-1},
\end{equation}
where $\omega_q$ the frequency of the $q$-polarized field.
\begin{equation}
    \hat{H}_{\text{NH}} = \hat{H}_e - \frac{i}{2}\sum_q \hat{L}_q^\dagger \hat{L}_q,
\end{equation}
where the inverse is defined in the excited-state basis. For each of the ${}^{87}$Rb D lines, transitions are modelled by the dipole operator
 \begin{equation}
        \hat{D}_q(F,F') =\sum_{m_F} C^{F,F'}_{m_F,q}\ket{F,m_F}\bra{F',m_F+q},
\end{equation}
and spontaneous emission is treated collectively within each polarization
\begin{equation}
        \hat{L}_q = \sqrt{\gamma}\sum_{F,F'}\hat{D}_{q}(F,F').
\end{equation}
Applying the formulae, the resulting effective super-operator couples states within the same hyperfine level, as well as causing transitions between the two. By selecting an operating regime where the population is confined to a single $F$ level, the dynamics are dominated by the elastic-scattering Hamiltonian
\begin{equation}
    \begin{split}
    \hat{H} =& \omega_\pi\hat{\Omega}_\pi^\dagger\hat{\Omega}_\pi + \omega_+\hat{\Omega}_+^\dagger\hat{\Omega}_+ + \omega_-\hat{\Omega}_-^\dagger\hat{\Omega}_-\\
    &+ \omega_{lS}\left(\hat{\Omega}_+^\dagger\hat{\Omega}_+-\hat{\Omega}_-^\dagger\hat{\Omega}_-\right)\hat{S}_z\\
    &+ \omega_{qS}\left(\hat{\Omega}_+^\dagger\hat{\Omega}_++\hat{\Omega}_-^\dagger\hat{\Omega}_--2\hat{\Omega}_\pi^\dagger\hat{\Omega}_\pi\right)\hat{S}_z^2 \\ 
    &+
    \lambda\left[\hat{S}_+\left(\hat{\Omega}_-^\dagger\hat{\Omega}_\pi+\hat{\Omega}_\pi^\dagger\hat{\Omega}_+\right)+\left(\hat{\Omega}_\pi^\dagger\hat{\Omega}_-+\hat{\Omega}_+^\dagger\hat{\Omega}_\pi\right)\hat{S}_-\right]\\
    &+
    \zeta\left[\left(\hat{\Omega}_+^\dagger-\hat{\Omega}_-^\dagger\right)\hat{\Omega}_\pi+\hat{\Omega}_\pi^\dagger\left(\hat{\Omega}_+-\hat{\Omega}_-\right)\right]\hat{Q}_{xz} \\
    &- i\zeta\left[\left(\hat{\Omega}_+^\dagger+\hat{\Omega}_-^\dagger\right)\hat{\Omega}_\pi-\hat{\Omega}_\pi^\dagger\left(\hat{\Omega}_+ +\hat{\Omega}_-\right)\right]\hat{Q}_{yz}\\
    &+
    \mathcal{T}(\hat{\Omega}_-^\dagger\hat{\Omega}_+\hat{S}_+^2 + \hat{S}_-^2\hat{\Omega}_+^\dagger\hat{\Omega}_-),
    \end{split}
    \label{GeneralHam}
\end{equation}
where $\hat{\Omega}_q$ is a $q$-polarized field operator. The first line indicates effective shifts in the fields of each polarization, while the second and third line are linear and quadratic AC Stark shifts, respectively. The remaining terms describe Raman transitions due to coupling between the atom and light fields.

The Hamiltonian \ref{GeneralHam} is a general form for an alkali atom coupled to polarized fields, and can now be tailored to the specific applications considered in this work. Assuming that any Zeeman shifts are small relative to the detunings considered, formulae for the effective parameters are readily derived, dependent only on the bare properties of the atom and the optical fields to which it couples.

\subsection{D1 Line}
The D1 line is driven by two auxiliary lasers, each of opposite circular-polarization, and coupled to a $\pi$-polarized cavity mode. This is incorporated into Hamiltonian (\ref{GeneralHam}), by explicitly writing the generic field operators as
\begin{equation}
    \hat{\Omega}_\pm\rightarrow\frac{\Omega_\pm}{2},~\hat{\Omega}_\pi\rightarrow g\hat{a}.
\end{equation}
Ignoring c-number terms, the remaining effective parameters are given below, for each ground-state hyperfine level.

\subsubsection{$F=1$ terms}
\begin{equation}
    \begin{split}
        \omega_\pi&=\frac{1}{3\Delta_{12'}},\\
        \omega_{lS}&=-\frac{1}{24}\left(\frac{1}{\Delta_{11'}}-\frac{5}{\Delta_{12'}}\right),\\
        \omega_{qS}&=-\frac{1}{24}\left(\frac{1}{\Delta_{11'}}-\frac{1}{\Delta_{12'}}\right),\\
        \lambda&=-\frac{1}{\sqrt{2}}\omega_{lS},\\
        \mathcal{T}&=\omega_{qS},\\
        \zeta&=\frac{1}{\sqrt{2}}\omega_{qS}
    \end{split}
\end{equation}

\subsubsection{$F=2$ terms}
\begin{equation}
    \begin{split}
        \omega_\pi&=\frac{1}{3\Delta_{21'}},\\
        \omega_{lS}&=-\frac{1}{24}\left(\frac{3}{\Delta_{21'}}+\frac{1}{\Delta_{22'}}\right),\\
        \omega_{qS}&=\frac{1}{24}\left(\frac{1}{\Delta_{21'}}-\frac{1}{\Delta_{22'}}\right),\\
        \lambda&=-\frac{1}{\sqrt{2}}\omega_{lS},\\
        \mathcal{T}&=\omega_{qS},\\
        \zeta&=\frac{1}{\sqrt{2}}\omega_{qS}.
    \end{split}
\end{equation}

The Hamiltonian (\ref{hamiltonianD1}) is recovered from these general terms, to give an effective evolution of the ground states due to D1 coupling. Explicitly, the parameters are related by:
\begin{equation}
\begin{split}
    \omega &= g^2\omega_\pi,\\
    \omega_0 &= \left(\frac{\Omega^2_+}{4}-\frac{\Omega^2_-}{4}\right)\omega_{lS},\\
    \lambda_\pm &= g\frac{\Omega_\pm}{2}\lambda,\\
    \zeta_\pm &= \mp g\left(\frac{\Omega_+}{2}\pm\frac{\Omega_-}{2}\right)\zeta,\\
    \tau &= \frac{\Omega_+\Omega_-}{4}\mathcal{T},\\
    \omega_q &= \left(\frac{\Omega_+^2}{4}+\frac{\Omega^2_-}{4}\right)\omega_{qS},\\
    \delta_q &= 2g^2\omega_{qS}.
    \end{split}
\end{equation}

\subsection{D2 Line}
The D2 is allowed to be driven by coherent lasers with any of the three polarization. This is described by replacing the general field operators in Hamiltonian (\ref{GeneralHam}) with
\begin{equation}
    \hat{\Omega}_q\rightarrow\frac{\Omega_q}{2}.
\end{equation}
Excluding constant terms, and again assuming that any Zeeman shifts are small relative to the transition detunings, the the effective parameters are given below for dynamics within each of the ground levels.
\subsubsection{$F=1$ terms}
\begin{equation}
    \begin{split}
        \omega_{lS}&=\frac{1}{48}\left(\frac{5}{\delta_{12'}}-\frac{5}{\delta_{11'}}-\frac{4}{\delta_{10'}}\right),\\
        \omega_{qS}&=\frac{1}{48}\left(\frac{1}{\delta_{12'}}-\frac{5}{\delta_{11'}}+\frac{4}{\delta_{10'}}\right),\\
        \lambda&=-\frac{1}{\sqrt{2}}\omega_{lS},\\
        \mathcal{T}&=\omega_{qS},\\
        \zeta&=\frac{1}{\sqrt{2}}\omega_{qS}.
    \end{split}
\end{equation}

\subsubsection{$F=2$ Terms}

\begin{equation}
    \begin{split}
        \omega_{lS}&=\frac{1}{240}\left(\frac{28}{\delta_{23'}}-\frac{5}{\delta_{22'}}-\frac{3}{\delta_{21'}}\right),\\
        \omega_{qS}&=\frac{1}{240}\left(\frac{4}{\delta_{23'}}-\frac{5}{\delta_{22'}}+\frac{1}{\delta_{21'}}\right),\\
        \lambda&=-\frac{1}{\sqrt{2}}\omega_{lS},\\
        \mathcal{T}&=\omega_{qS},\\
        \zeta&=\frac{1}{\sqrt{2}}\omega_{qS}.
    \end{split}
\end{equation}

The Hamiltonian (\ref{hamiltonianD2}) is recovered from these general terms, where the effective parameters for ground-state evolution due to D2 coupling are given by:
\begin{equation}
\begin{split}
    \omega_0' &= \left(\frac{\Omega^2_+}{4}-\frac{\Omega^2_-}{4}\right)\omega_{lS},\\
    \lambda' &= \frac{\Omega_\pi}{2}\left(\frac{\Omega_+}{2}+\frac{\Omega_+}{2}\right)\lambda,\\
    \zeta'_\pm &= \mp \frac{\Omega_\pi}{2}\left(\frac{\Omega_+}{2}\pm\frac{\Omega_-}{2}\right)\zeta,\\
    \tau' &= \frac{\Omega_+\Omega_-}{4}\mathcal{T},\\
    \omega_q' &= \left(\frac{\Omega_+^2}{4}+\frac{\Omega^2_-}{4}-2\frac{\Omega^2_\pi}{4}\right)\omega_{qS}.
    \end{split}
\end{equation}

\end{document}